%% file: main.tex
\pdfoutput=1

\documentclass[11pt]{article}

\usepackage[]{ACL2023}

\usepackage{times}
\usepackage{latexsym}
\usepackage{hyperref}
\usepackage{graphicx}
\usepackage{verbatim}
\usepackage{booktabs}
\usepackage{float}

\definecolor{backcolour}{rgb}{0.95,0.95,0.95}
\definecolor{codegreen}{rgb}{0,0.5,0}
\definecolor{codegray}{rgb}{0.5,0.5,0.5}
\definecolor{codepurple}{rgb}{0.58,0,0.82}

\usepackage[T1]{fontenc}

\usepackage[utf8]{inputenc}

\usepackage{microtype}

\usepackage{inconsolata}

\usepackage{multicol} 

\usepackage{inconsolata}
\usepackage{listings}
\usepackage{color}
\usepackage{algorithm} 
\usepackage{algpseudocode} 
\usepackage{graphicx} 
\usepackage{float}    
\usepackage{pifont}
\usepackage{booktabs} 
\usepackage{tabularx} 
\usepackage{geometry} 

\usepackage{amsmath} 

\usepackage{times}
\usepackage{microtype}
\usepackage{epsfig}
\usepackage{caption}
\usepackage{float}
\usepackage{placeins}
\usepackage{color, colortbl}
\usepackage{stfloats}
\usepackage{enumitem}
\usepackage{tabularx}
\usepackage{xstring}
\usepackage{multirow}
\usepackage{xspace}
\usepackage{url}
\usepackage{subcaption}
\usepackage{xcolor}
\usepackage[hang,flushmargin]{footmisc}

\usepackage{natbib}
\setcitestyle{numbers,open={[},close={]}}

\lstdefinestyle{mystyle}{
    backgroundcolor=\color{backcolour},   
    commentstyle=\color{codegreen},
    keywordstyle=\color{magenta},
    numberstyle=\tiny\color{codegray},
    stringstyle=\color{codepurple},
    basicstyle=\footnotesize,
    breakatwhitespace=false,         
    breaklines=true,                 
    captionpos=b,                    
    keepspaces=true,                 
    numbers=left,                    
    numbersep=5pt,                  
    showspaces=false,                
    showstringspaces=false,
    showtabs=false,                  
    tabsize=2
}

\newcommand{\thankswithlink}[1]{%
  \begingroup
  \renewcommand\thefootnote{}
  \footnote{%
    \noindent
    \fontsize{9.6}{10}\selectfont 
    \href{#1}{#1}%
  }%
  \addtocounter{footnote}{-1}
  \endgroup
}

\title{Code Sharing In Prediction Model Research: A Scoping Review}

\author{
    Thomas Sounack$^{1}$
  \\\And
   Raffaele Giancotti$^{2}$
   \\\And
   Catherine A. Gao$^{3}$
  \\\AND
  Lasai Barreñada$^{4}$
  \\\And
  Hyeonhoon Lee$^{5,6}$
  \\\And
  Hyung-Chul Lee$^{5,6}$
  \\\And
  Leo Anthony Celi$^{7,8}$
   \\\AND
  Karel G.M. Moons $^{9}$
  \\\And
   Gary S. Collins$^{10,11}$
    \\\And
   Charlotta Lindvall$^{1,12,\dagger}$
  \\\And
    Tom Pollard$^{7,8,\dagger}$
}

\affil{
   $^1$ Dana-Farber Cancer Institute \hspace{0.3cm}
   $^2$ University of Calabria \hspace{0.3cm}
   $^3$ Northwestern Feinberg School of Medicine \\\vspace{1mm}
   $^4$ KU Leuven \hspace{0.3cm}
   $^5$ Seoul National University Hospital \hspace{0.3cm}
   $^6$ Seoul National University College of Medicine \\\vspace{1mm}
   $^7$ Massachusetts Institute of Technology \hspace{0.2cm}
   $^8$ Harvard T.H. Chan School of Public Health \hspace{0.2cm} 
   $^9$ UMC Utrecht
   \\\vspace{1mm}
   $^{10}$ University of Birmingham
   \hspace{0.3cm}
   $^{11}$ University Hospitals Birmingham NHS Foundation Trust
    \\\vspace{1mm}
    $^{12}$ Harvard Medical School
     \\\vspace{5mm}
    $^\dagger$ Co-senior authors
}

\begin{document}

\maketitle

\thankswithlink{$^1$https://github.com/thomas-sounack/TRIPOD-Code}

\begin{abstract}

\input{Sections/0_abstract}

\end{abstract}

\input{Sections/1_introduction}
\input{Sections/2_methods}
\input{Sections/3_results}
\input{Sections/4_discussion}
\input{Sections/5_conclusion}
\input{Sections/6_others}

\bibliographystyle{unsrtnat}
\bibliography{custom}



\appendix

\input{Sections/99_appendix}

\end{document}

%% file: Sections/0_abstract.tex
\textbf{Background} Analytical code is essential for reproducing diagnostic and prognostic prediction model research, yet code availability in the published literature remains limited. While the TRIPOD statements set standards for reporting prediction model methods, they do not define explicit standards for repository structure and documentation. This review quantifies current code-sharing practices to inform the development of TRIPOD-Code, a TRIPOD extension reporting guideline focused on code sharing.

\textbf{Methods} We conducted a scoping review of PubMed-indexed articles citing TRIPOD or TRIPOD+AI as of Aug 11, 2025, restricted to studies retrievable via the PubMed Central Open Access API. Eligible studies developed, updated, or validated multivariable prediction models. A large language model (LLM)-assisted pipeline (GPT-5.2 2025-12-11) was developed to screen articles and extract code availability statements and repository links. Repositories were downloaded using a code retrieval utility and assessed with the same LLM against 14 predefined reproducibility-related features, including the presence of a README file, versioned dependencies, and modularity. Our code is made publicly available$^1$.

\textbf{Findings} Among 3,967 eligible articles, 12.2\% (n=482) included code sharing statements. Code sharing increased over time, reaching 15.8\% in 2025, and was higher among TRIPOD+AI-citing studies than TRIPOD-citing studies (29.7\% vs. 11.8\% in 2025, excluding articles citing both). Sharing prevalence varied widely by journal and country, with higher sharing observed in journals that discuss code availability in their submission guidelines. Of 447 extracted repository links, 83.4\% were hosted on GitHub. Repository assessment showed substantial heterogeneity in reproducibility features: most repositories contained a README file (80.5\%), but fewer specified dependencies (37.6\%; version-constrained 21.6\%) or were modular (42.4\%). Data or sample datasets were included in 37.1\% of repositories, and among the 286 repositories that included stochastic components, 64.3\% fixed random seeds.

\textbf{Interpretation} In prediction model research, code sharing remains relatively uncommon, and when shared, often falls short of being reusable. These findings provide an empirical baseline for the TRIPOD-Code extension and underscore the need for clearer expectations beyond code availability, including documentation, dependency specification, licensing, and executable structure.

%% file: Sections/1_introduction.tex
\section{Introduction}

Development and evaluation of diagnostic and prognostic models are inherently computational. The analytical code used to, e.g., preprocess data, develop the prediction modeling approach, select or tune model parameters, and evaluate performance is essential for methodological assessment and reproduction of results. Yet access to the code underlying published prediction models remains uncommon \cite{collinsOpenSciencePractices2024}.

Concerns about computational transparency in scientific research are longstanding, and recent initiatives have sought to strengthen reporting and sharing practices. The FAIR principles have outlined standards for reusable digital research artifacts \cite{wilkinsonFAIRGuidingPrinciples2016}; the EQUATOR Network has advanced reporting guidelines to enhance methodological transparency \cite{simeraTransparentAccurateReporting2010}; and journals and funders increasingly require data and code availability statements to promote reproducibility \cite{DoesYourCode2018, bloomDataAccessOpen2014, colavizzaCitationAdvantageLinking2020}.

Within clinical prediction research, the TRIPOD statement and its update in 2024, TRIPOD+AI (www.tripod-statement.org) \cite{collinsTransparentReportingMultivariable2015, moonsTransparentReportingMultivariable2015, collinsTRIPOD+AIStatementUpdated2024}, were developed to improve the completeness and clarity of model reporting. These guidelines have helped promote methodological quality, external validation, and calibration of prediction model research \cite{zamanipoornajafabadiTRIPODStatementPreliminary2020}. However, although TRIPOD+AI encourages sharing of analytical code, it does not explicitly specify standards for repository structure, documentation, or reproducibility. Consequently, reporting of code availability is likely to be heterogeneous \cite{streiberImprovingReproducibilityData2025}, although the actual extent and quality of code sharing in prediction model research have not yet been studied.

Evidence from broader biomedical literature indicates that code sharing is infrequent and that declared code availability often does not correspond to functional accessibility \cite{hamiltonPrevalencePredictorsData2023, sharmaAnalyticalCodeSharing2024}. In a large-scale attempt to rerun Jupyter notebooks linked to biomedical articles in PubMed Central, prior work identified thousands of Python Jupyter notebooks but found that only 5.3\% could be run successfully end-to-end \cite{samuelComputationalReproducibilityJupyter2024}. Manual review of code repositories reports similar gaps: among articles accepted at the Medical Imaging with Deep Learning (MIDL) conference, only 22\% had a public repository judged “repeatable” under predefined reproducibility criteria \cite{simkoReproducibilityMethodsMedical2024}, and in a study evaluating 160 deep learning computational pathology articles, only about one quarter were found to make code publicly available \cite{wagnerBuiltLastReproducibility2024}.

However, these investigations focused on specific code formats (e.g., Jupyter notebooks) or assessed code documentation through a labor-intensive manual review that does not scale to large, diverse corpora. Prior work has shown the potential of large language models (LLMs) to facilitate large-scale reviews \cite{chenLLMassistedSystematicReview2026, delgado-chavesTransformingLiteratureScreening2025}, but to our knowledge, no study has assessed code sharing and repository documentation practices in prediction model research, applied LLMs to jointly extract code-sharing statements from articles, or characterized repository features at scale. Establishing this empirical baseline is necessary to understand the current practice of code sharing in prediction model research and to inform the development of more precise reporting standards.

In response to this need, we are developing the TRIPOD-Code extension to provide structured guidance on code availability and computational reproducibility in prediction model research \cite{pollardProtocolDevelopmentReporting2026}. The present study represents Stage 1 of this project: a large-scale scoping review of studies citing TRIPOD or TRIPOD+AI, undertaken to quantify code availability and evaluate the characteristics of shared repositories.

%% file: Sections/2_methods.tex
\section{Methods}

We conducted a scoping review to characterize code sharing practices for papers developing or evaluating multivariable prediction models in healthcare. The review addresses two key questions: (1) \textit{What proportion of multivariable prediction model studies that cite the TRIPOD or TRIPOD+AI statements report on the availability of analytical code?} (2) \textit{Among studies providing accessible code, what are the structural and documentation characteristics of the code?} The review was conducted in accordance with the PRISMA extension for scoping reviews \cite{triccoPRISMAExtensionScoping2018} (\autoref{prisma-checklist}). The protocol for this review was registered (INPLASY202620080) \cite{sounackCodeSharingPrediction2026}.

\subsection{Search strategy and selection criteria}

The cohort comprised all PubMed-indexed articles that cited the TRIPOD or TRIPOD+AI statements (www.tripod-statement.org) as of August 11th, 2025. Focusing on TRIPOD-citing studies provides a defined and policy-relevant subset of prediction model research in which reporting guidance is already acknowledged. As such, this sampling frame represents a conservative estimate of code sharing practices among studies attentive to reporting standards.

Due to simultaneous publication across different journals, TRIPOD and TRIPOD+AI have multiple entries in PubMed (12 and 2 entries, respectively). For each entry, citation lists were downloaded from the PubMed “Cited by” section and aggregated programmatically. Duplicate records were removed. We included primary research articles that developed, validated, or updated a multivariable prediction model using statistical or machine learning methods, consistent with definitions in the TRIPOD-Code protocol \cite{pollardProtocolDevelopmentReporting2026}. To enable reproducibility without subscription barriers, we restricted inclusion to articles retrievable through the PubMed Central Open Access API \cite{PMCOpenAccess2003}. No additional exclusion criteria were applied at this stage.

\subsection{Data analysis}

Data analysis comprised two components: (1) article-level screening and metadata extraction and (2) characterization of associated code repositories. An LLM-assisted pipeline was implemented using a predefined structured output schema (Appendices \ref{prompt-articles} and \ref{prompt-repositories}). All model outputs were validated against human-annotated subsets. The codebook used by the annotators is provided in \autoref{annotation-codebook}.

\subsubsection{Article screening and metadata extraction}

Full-text articles were retrieved using the PubMed Central Open Access API. Figures, tables and appendices were excluded from the extraction. Eligibility screening and metadata extraction were performed using an LLM (GPT-5.2 2025-12-11). The LLM was prompted using a structured schema to determine eligibility, identify code availability statements, extract repository links, and record the country of the first author’s institution (\autoref{prompt-articles}). 

To evaluate the performance of the automated pipeline, two independent reviewers (TS, RG) manually annotated 500 randomly selected articles. Disagreements were resolved through group discussion involving both annotators. Performance of the pipeline was then evaluated against the human labels. The LLM achieved a weighted F1 score of 0.97 on the selection criteria and a 92.3\% accuracy on repository link retrieval. Full evaluation metrics and error analysis are provided in \autoref{eval-articles}.

Repositories identified during article screening were retrieved using a custom repository retrieval utility (described below). The utility supported links from GitHub, GitLab, Gitee, Zenodo, Figshare, Open Science Framework (OSF), and Digital Object Identifiers (DOIs) resolving to these providers. Repositories were retrieved as of the date of analysis and processed from their default branch. An article was considered to be sharing code if the retrieval utility successfully downloaded repository contents and detected at least one non-empty source code file. We also counted articles that stated code was available in supplementary materials, or that linked to a repository hosted on a provider unsupported by our retrieval utility. For these latter cases, we could not independently verify accessibility or content, but we included them to avoid underestimating reported code availability. Code availability was analyzed by publication year, journal, and country of first author affiliation. The location and wording of code sharing statements were also categorized.

Once code repositories were identified, they were explored and characterized. A repository retrieval utility was developed to compile each repository’s content into a structured text file. These files were subsequently processed by an LLM to extract code properties.

\subsubsection{Code repository characterization}

Repositories retrieved during article screening were subsequently characterized to assess code documentation, structure, and features relevant to reproducibility. A custom utility compiled repository content and metadata into a textual representation according to predefined inclusion rules. README files and code files were included in full. Large binary files (e.g., datasets, model checkpoints, and compiled artifacts) were excluded. Other file types were truncated to a maximum of 3,000 tokens to avoid exceeding the model context window.

Characterization was performed using an LLM (GPT-5.2 2025-12-11) prompted with a structured schema defining 14 repository features prior determined by the TRIPOD-Code executive committee (\autoref{prompt-repositories}). These features captured elements relevant to computational reproducibility, including the presence of documentation (e.g., README), licensing information, specification of dependencies, test frameworks, and indications of version control. In addition to the structured feature extraction, the LLM generated brief free-text summaries of notable strengths and weaknesses of each repository for qualitative review.

To evaluate performance, two independent reviewers (TS, RG) annotated all retrieved repositories (n=35) according to the same predefined schema. Disagreements were resolved through group discussion involving both annotators. Model performance was assessed against the human labels, yielding a weighted F1 score of 0.83 over all features. Performance was highest for objective features such as the presence of a README file (F1=1.0), LICENSE file (F1=1.0), and tests (F1=1.0), and lower for more subjective criteria such as adequacy of documentation (F1=0.71). Detailed performance metrics and error analysis are provided in \autoref{eval-repos}. Repository characteristics were summarized descriptively at the cohort level and analyzed in relation to the journal of publication and year.

%% file: Sections/3_results.tex
\section{Results}

\begin{figure}[H]
    \centering
    \includegraphics[width=0.8\textwidth]{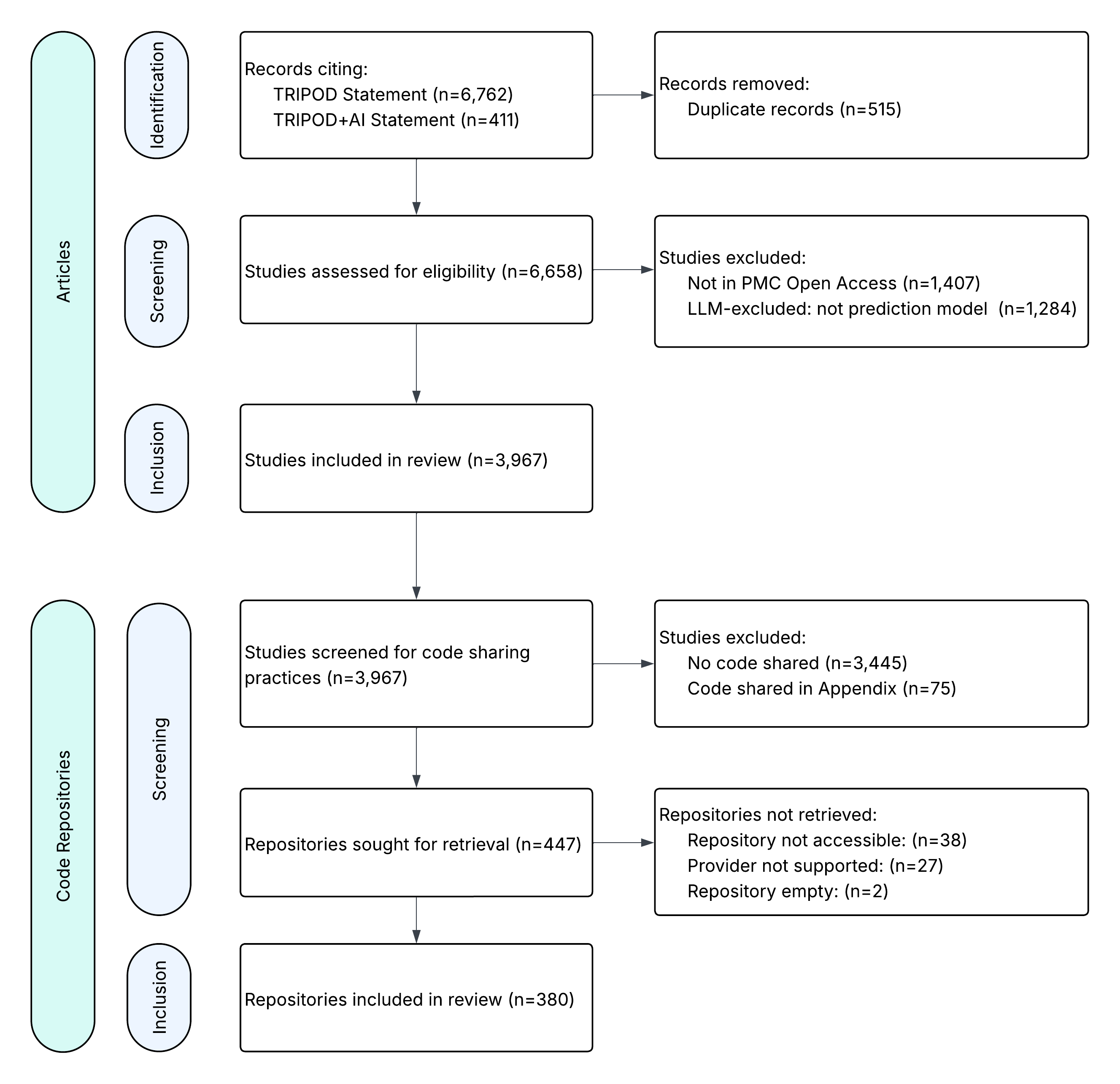}
    \captionsetup{width=0.9\textwidth}
    \caption{Modified PRISMA Flow diagram. We determined that an article was sharing code (n=482) if: it shared code in the Appendix (n=75), its repository provider was not supported (n=27) or its repository was included in the review (n=380).}
    \label{fig:prisma-flow-diagram}
\end{figure}

We identified 6,762 PubMed articles citing the TRIPOD Statement and 411 citing the TRIPOD+AI Statement as of August 11th, 2025 (\autoref{fig:prisma-flow-diagram}). After removing 515 duplicate entries, the initial cohort comprised 6,658 unique articles with associated PubMed metadata. Of these, 1,407 articles were not retrievable through the PMC Open Access API, and 1,284 articles did not develop, update, or validate a multivariable prediction model using a statistical or machine learning technique, leaving 3,967 papers that met our selection criteria (see \autoref{csv-articles-selected}).

\subsection{Code sharing}

\begin{figure}[htbp]
    \centering
    \includegraphics[width=0.85\textwidth]{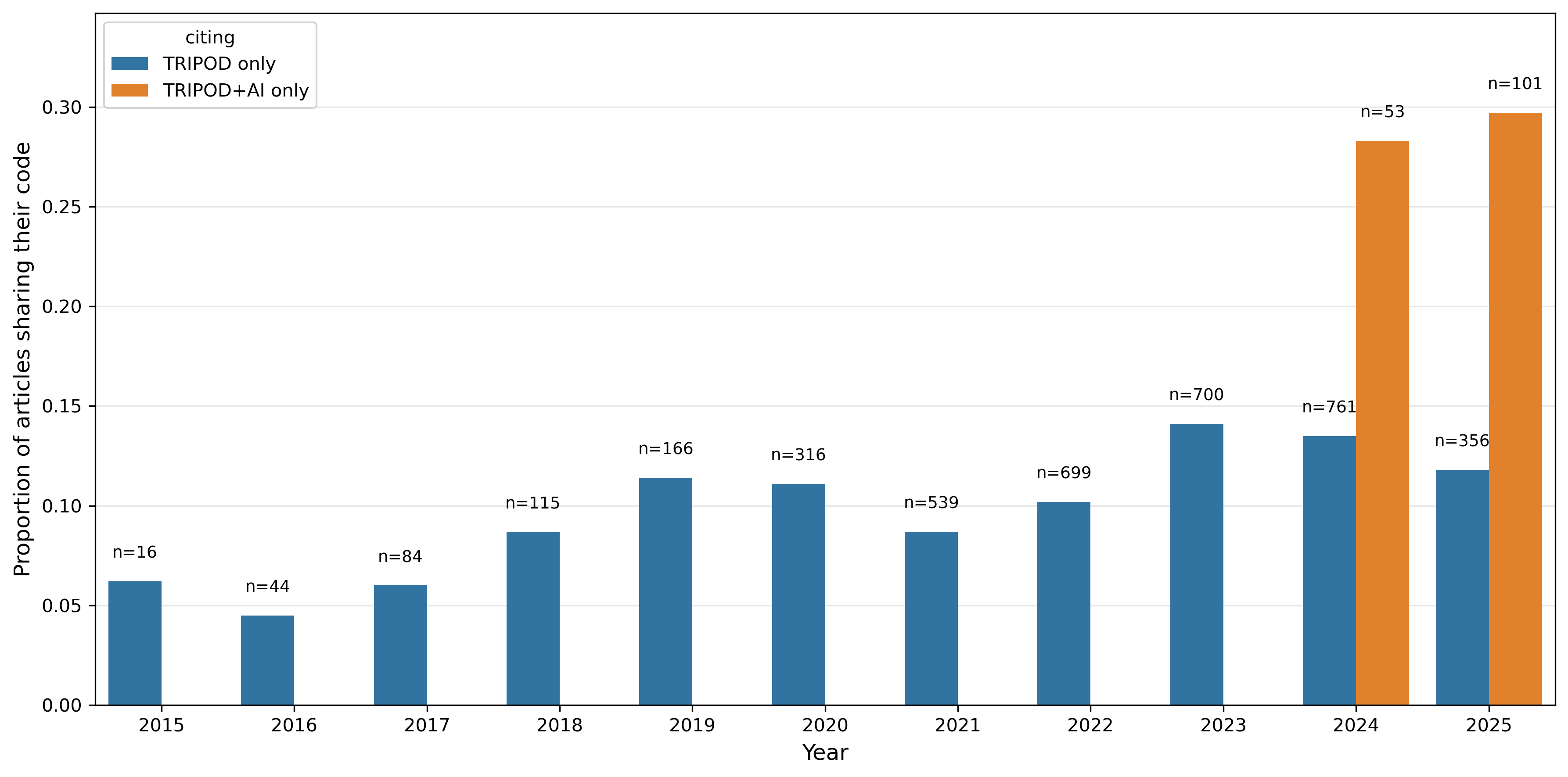}
    \captionsetup{width=0.9\textwidth}
    \caption{Proportion of articles sharing their code by year, for the article cohort (n=3,967). Articles that only cited the TRIPOD statement are displayed in blue, and articles that only cited the TRIPOD+AI statement are displayed in orange. 17 articles cited both statements, of which 3 reported their code, and are excluded from this figure.}
    \label{fig:code-sharing-over-time}
\end{figure}

Among the 3,967 articles included in the review, 12.2\% (n=482) shared their code. Code was most frequently shared via a repository link, while 75 articles reported code availability in an appendix or supplementary material. Code sharing increased over time across the study period (\autoref{fig:code-sharing-over-time}). The proportion of articles sharing code rose from 6.3\% in 2015 (1/16) and 4.5\% in 2016 (2/44) to 14.5\% in 2024 (119/820) and 15.8\% in 2025 (74/468). Articles citing the TRIPOD+AI statement showed a markedly higher code sharing prevalence (29.2\%) compared with those citing TRIPOD (11.4\%). In 2025, 29.7\% (30/101) of the articles citing TRIPOD+AI shared code, compared to 11.8\% (42/356) for articles citing TRIPOD in that same year (excluding 11 articles citing both).

\begin{figure}[H]
    \centering
    \includegraphics[width=0.85\textwidth]{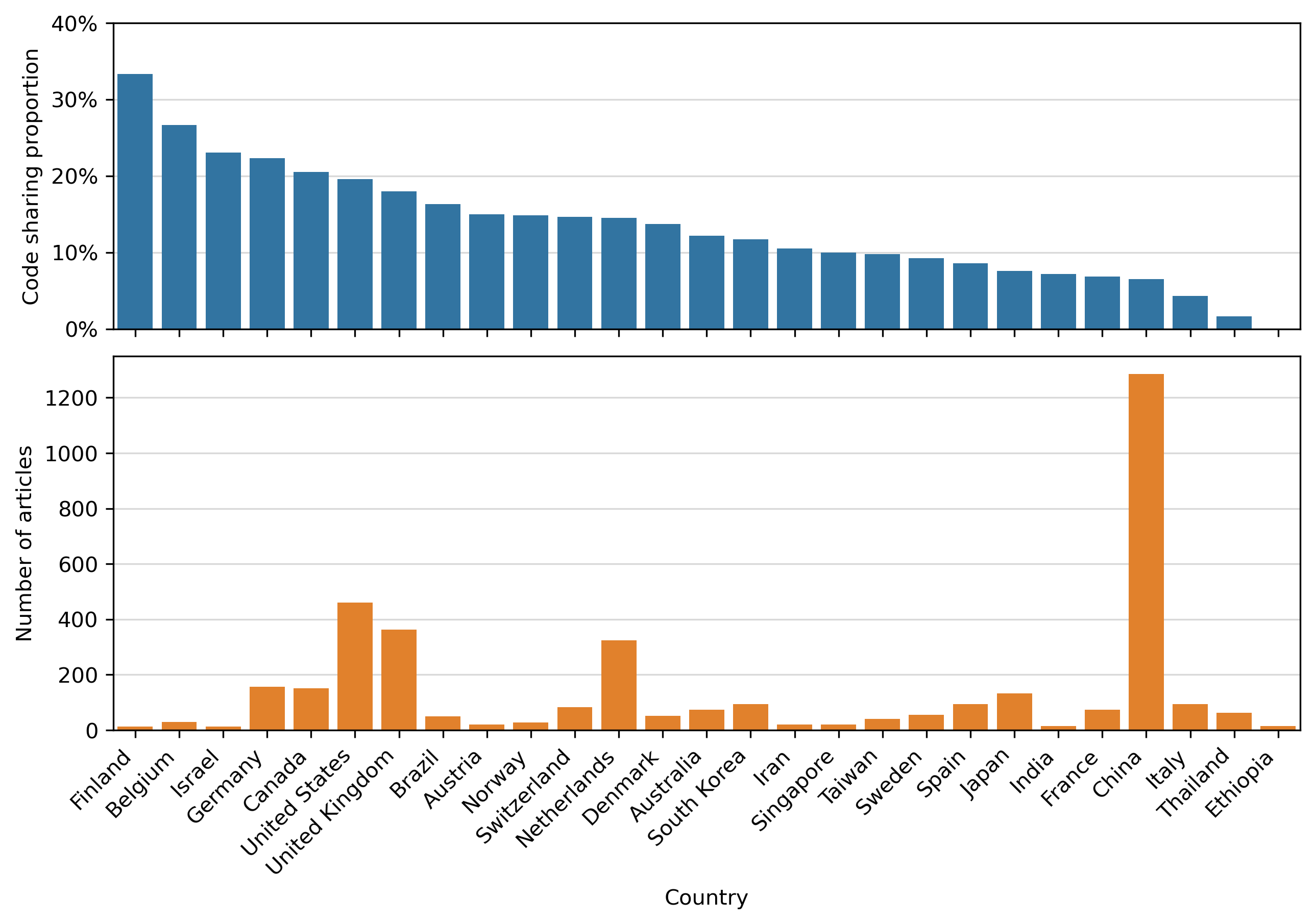}
    \captionsetup{width=0.9\textwidth}
    \caption{(Top) Proportion of articles sharing their code by country and (Bottom) number of articles by country, for the article cohort (n=3,967). Countries with more than 10 articles in the articles cohort are included.}
    \label{fig:code-sharing-countries}
\end{figure}

China accounted for the largest share of publications (32.4\%), followed by the United States (11.6\%) and the United Kingdom (9.1\%). Code sharing prevalence varied across countries, with the highest proportions observed in Finland (33.3\%), Belgium (26.7\%), and Israel (23.1\%) (\autoref{fig:code-sharing-countries}).

\begin{figure}[H]
    \centering
    \includegraphics[width=0.95\textwidth]{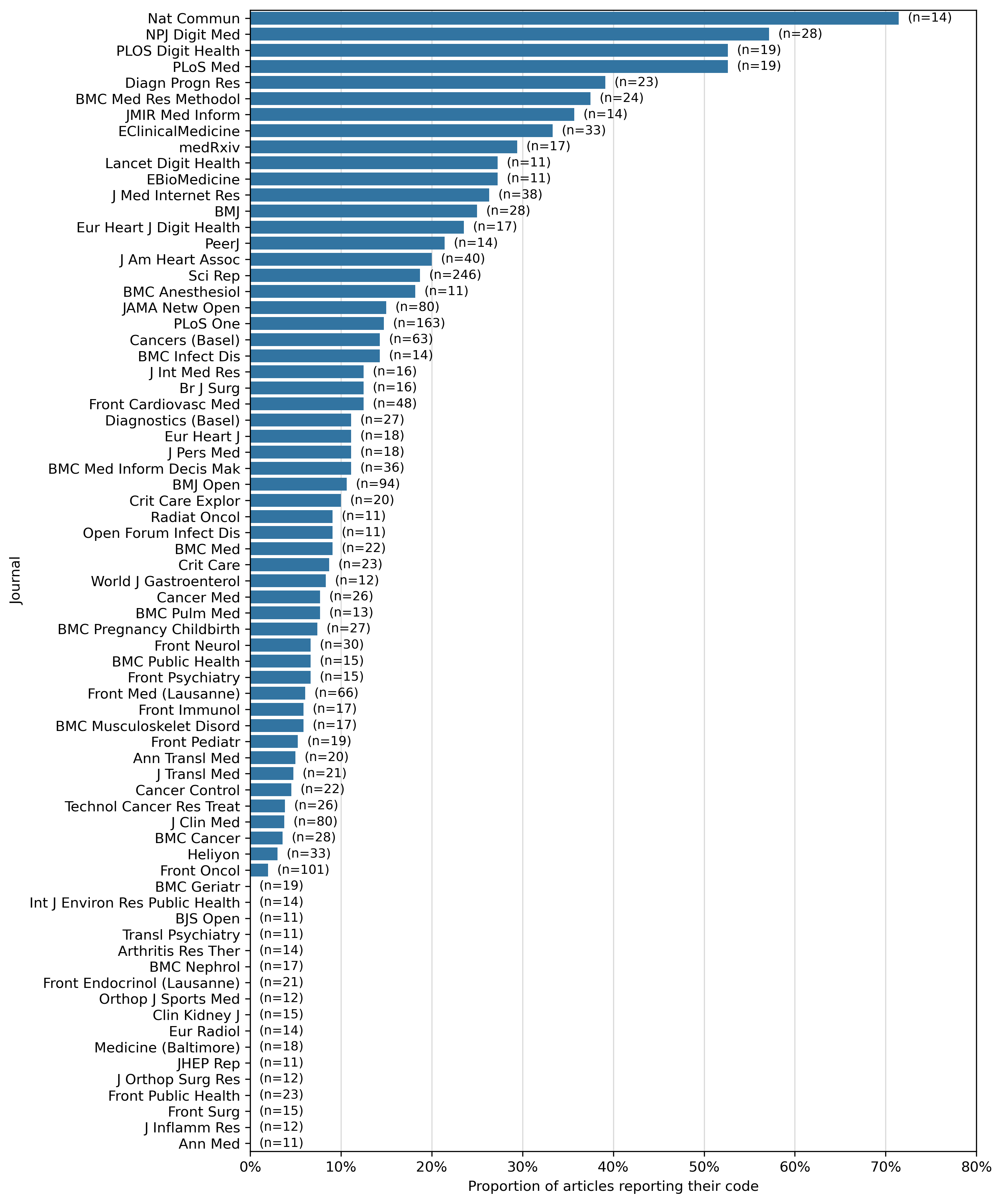}
    \captionsetup{width=0.9\textwidth}
    \caption{Proportion of articles reporting their code by journal, for the article cohort (n=3,967). Journals with more than 10 articles in the cohort are included.}
    \label{fig:code-sharing-journals}
\end{figure}

Code sharing prevalence varied by journal (\autoref{fig:code-sharing-journals}). Among journals with more than 10 articles included in the review, 17 out of 71 had a code sharing prevalence of 0\%, while the highest-scoring journals, \textit{Nature Communications} (n=14), \textit{npj Digital Medicine} (n=28), \textit{PLOS Digital Health} (n=19) and \textit{PLOS Medicine} (n=19) scored 71.4\%, 57.1\%, 52.6\% and 52.6\%, respectively.

Among papers providing a repository link, code availability statements most frequently appeared in the Data Availability section (n=232), followed by the Methods section (n=211) and the supplementary materials (n=84). Most articles included a single code sharing statement (68.6\%), while 29.1\% included two statements and 2.1\% included three.

We analyzed the phrasing of code sharing statements to assess linguistic consistency. The statements were highly heterogeneous; most phrasings (94.4\%) occurred only once. Statements that appeared more than once are listed in \autoref{code-sharing-sentences}. We also observed substantial variation in content. Some statements referred only to “code”, whereas others specified its purpose, such as model development and evaluation, or data preprocessing and analysis. A few explicitly mentioned that the code was open source. Most statements named the repository platform before providing the link. We noted that one journal, \textit{npj Digital Medicine}, provided example code availability statements in its submission guidelines. These example statements were not repeatedly observed.

\subsection{Repository characteristics}

Of 447 repository links identified, 27 links pointed to unsupported providers and 38 could not be resolved despite matching a supported provider. Unresolved links typically corresponded to private repositories, malformed URLs, or links to user profiles rather than to specific repositories. Two accessible repositories contained no source code, leaving a total of 380 repositories to be characterized. We found that 83.4\% of repositories were hosted on GitHub, with smaller proportions hosted on OSF (3.6\%), Zenodo (3.4\%), and GitLab (1.6\%).

\begin{figure}[H]
    \centering
    \includegraphics[width=0.9\textwidth]{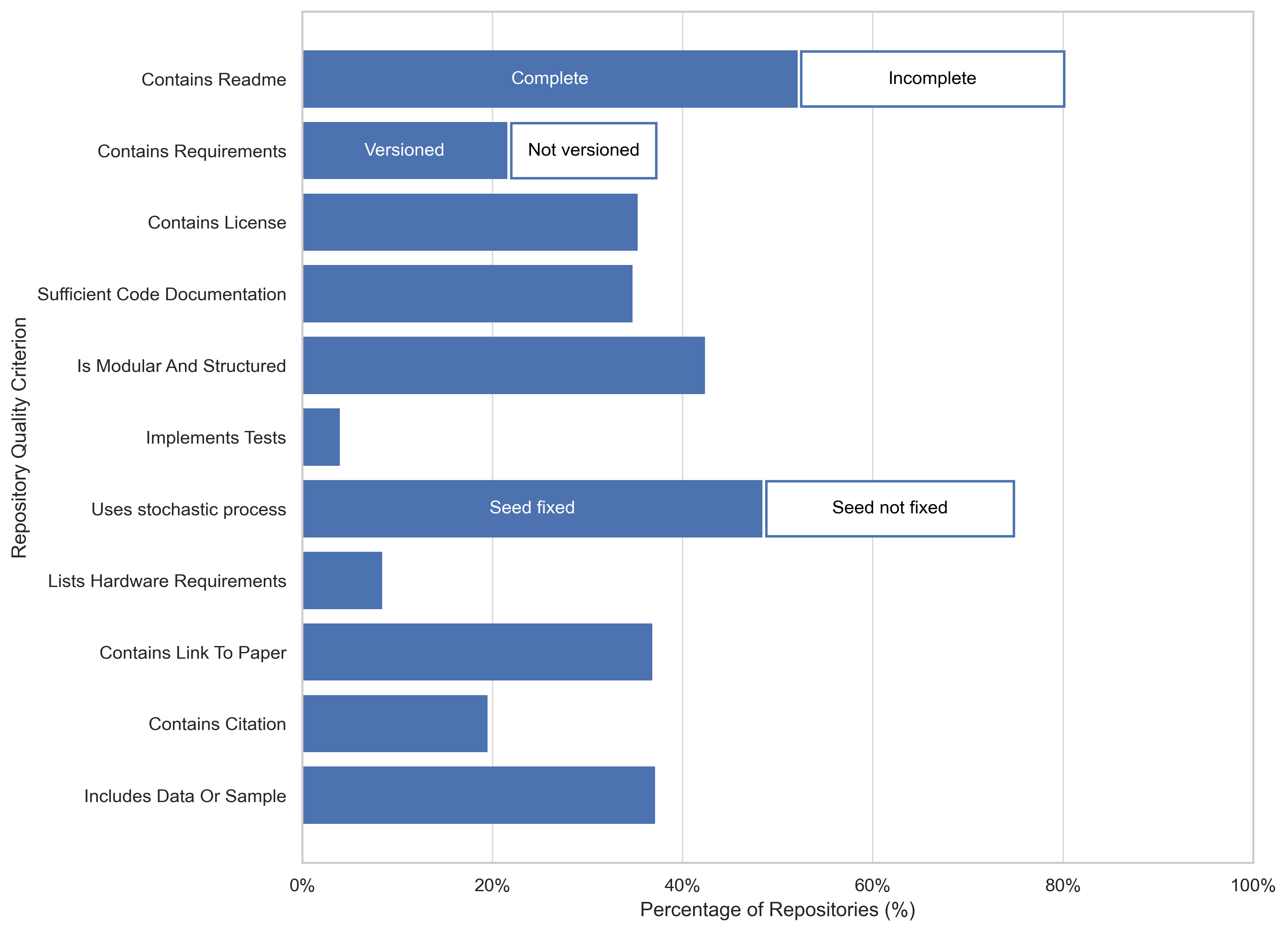}
    \captionsetup{width=0.9\textwidth}
    \caption{Repository characteristics – quality criteria against percentage of repositories, for the repository cohort (n=380). README files that provide an overview of the purpose and expected outputs of their code repository are shown as Complete, while those that do not are shown as Incomplete.}
    \label{fig:repo-practices}
\end{figure}

Documentation practices varied (\autoref{fig:repo-practices}). Most repositories (80.5\%) contained a README file, however only 52.1\% provided a README that described the repository’s purpose and expected outputs. In terms of design, 42.4\% were classified as modular and structured. Software dependencies were specified in 37.6\% of repositories, with versioned dependency constraints in 21.6\%. Licensing information was present in 35.3\% of repositories, a link to the associated publication in 36.8\%, and formal citation information in 19.5\%. Tests were rarely implemented (3.9\%), and hardware requirements were documented in 8.4\% of cases. Data or sample datasets were included in 37.1\% of repositories. Among the 286 repositories that included stochastic components, 64.3\% fixed random seeds.

\begin{figure}[H]
    \centering
    \includegraphics[width=0.95\textwidth]{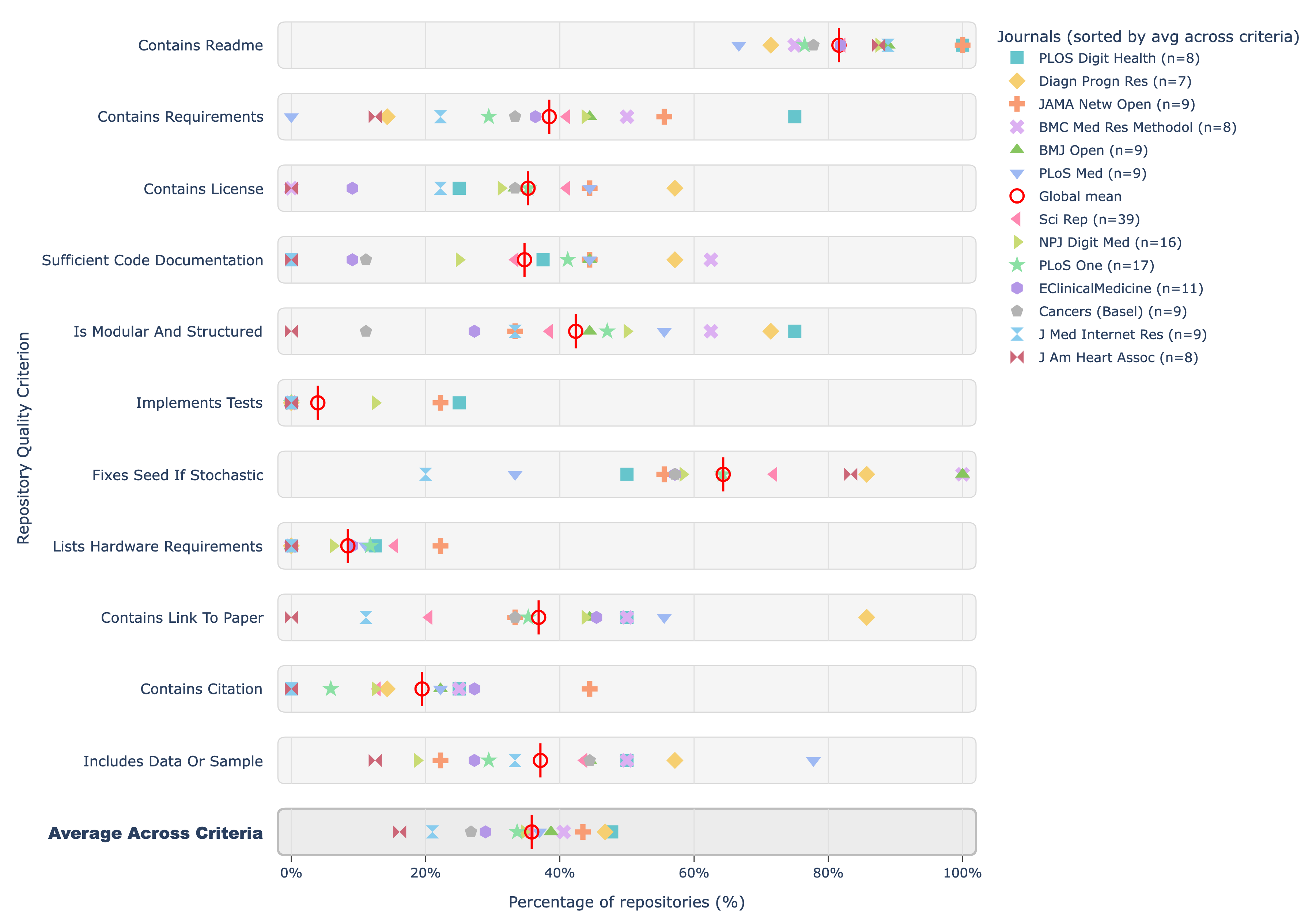}
    \captionsetup{width=0.9\textwidth}
    \caption{Repository characteristics by journal, for the repository cohort (n=380), for journals with more than 5 repositories. The legend displays journals sorted by descending average across criteria.}
    \label{fig:repo-practices-by-journal}
\end{figure}

Repository characteristics varied across journals (\autoref{fig:repo-practices-by-journal}). No journal ranked consistently highest or lowest across all criteria. Aggregated across criteria, \textit{PLOS Digital Health} had the highest global average, scoring 11.9\% above the global average across our criteria. In contrast, the lowest ranking journal scored 0\% on seven of the criteria. 

\begin{figure}[H]
    \centering
    \includegraphics[width=0.85\textwidth]{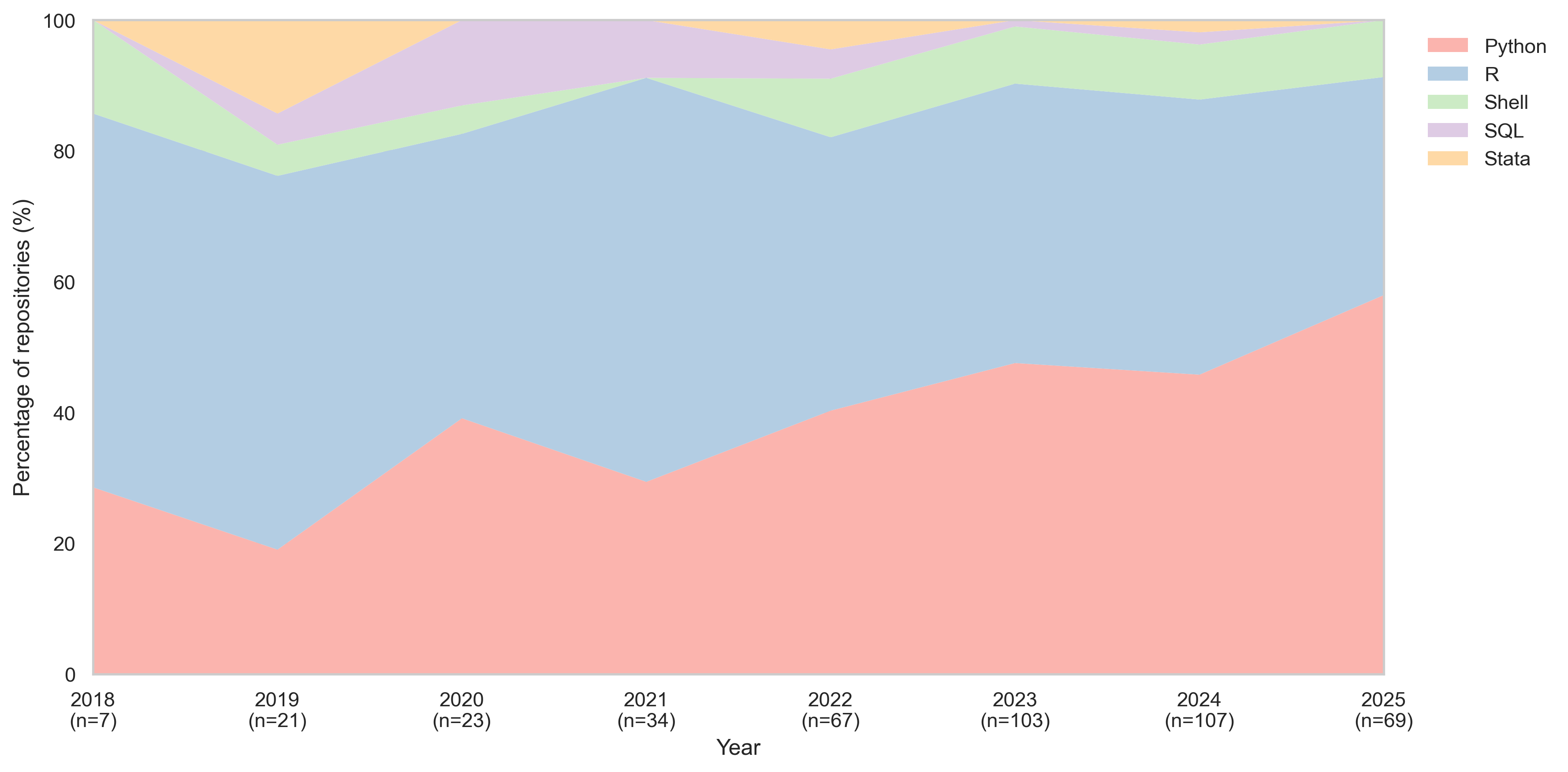}
    \captionsetup{width=0.9\textwidth}
    \caption{Annual distribution of programming languages across repositories, for the repository cohort (n=380) (languages appearing in >5 repositories overall). Years with less than 5 repositories were excluded.}
    \label{fig:language-distrib}
\end{figure}

Python and R were the most common programming languages, appearing in 190 and 189 repositories, respectively. Shell scripts (n=33) and SQL (n=13) were the next most frequent. Over time (\autoref{fig:language-distrib}), Python’s share increased steadily, overtaking R in 2023 as the most common language and accounting for more than half of all programming languages in 2025. The full language distribution is provided in \autoref{language-repo-distribution}.

%% file: Sections/4_discussion.tex
\section{Discussion}

This large-scale scoping review provides a repository-level assessment of code sharing practices among prediction model studies citing TRIPOD or TRIPOD+AI. Several findings stand out. First, code sharing remains relatively uncommon in prediction model research, with only 12.2\% of studies reporting accessible code. Second, while the availability of code has increased over time, substantial heterogeneity persists across journals and countries. Third, direct examination of repositories revealed that code availability does not consistently translate into reproducibility, with many repositories lacking features associated with reuse, such as documentation, dependency specification, and licensing. Availability appears to be a necessary but insufficient condition for computational reproducibility.

Variation across journals was pronounced, likely reflecting differences in journal-level guidance and expectations. This heterogeneity could also be attributed to differences in disciplinary backgrounds of contributing authors, as some journals publish more clinically oriented work whereas others more closely align with computational research. The journals with the highest proportion of code sharing - \textit{Nature Communications}, \textit{npj Digital Medicine}, \textit{PLOS Digital Health} and \textit{PLOS Medicine} (\textit{ex aequo}) - all provide guidance on code sharing. \textit{npj Digital Medicine} offers direct instructions on implementing a Code Availability section and states that the Editors “reserve the right to decline the manuscript if important code is unavailable” \cite{ReportingStandardsAvailability}. \textit{PLOS Digital Health} and \textit{PLOS Medicine} indicate that they expect researchers will make software available without restrictions upon publication of the work, and that it should “remain usable over time regardless of versions or upgrades” \cite{MaterialsSoftwareCode, MaterialsSoftwareCodea}. Importantly, higher rates of code sharing did not necessarily correspond to stronger repository practices, suggesting that mandating deposition alone may not be sufficient to ensure computational reproducibility. 

The contrast between studies citing TRIPOD and those citing TRIPOD+AI is notable. The proportion of articles citing TRIPOD+AI was nearly three times higher than that of articles citing TRIPOD. TRIPOD+AI includes an explicit item encouraging code availability (item 18f), whereas the original TRIPOD checklist does not explicitly reference code sharing - although it was discussed in the accompanying Explanation and Elaboration paper. Although differences in study type or journal distribution may partially explain this pattern, the finding aligns with broader literature indicating that structured reporting guidelines can shape author behavior \cite{aghaImpactMandatoryImplementation2016, panicEvaluationEndorsementPreferred2013, leclercqMetaanalysesIndexedPsycINFO2019}.

Beyond the availability of code, we found substantial variation in documentation and reproducibility features across repositories. While most repositories contained a README file, fewer than half demonstrated modular organization or explicit dependency specification. Version-constrained dependencies, formal licenses, test scripts, and citation metadata were also infrequent. Issues such as hard-coded file paths and reliance upon loosely referenced external utilities were found during manual review. These findings suggest that in many cases, repositories serve as an archival artifact rather than fully reusable computational resources. At the same time, we identified many examples of strong practice, including step-by-step instructions for reproduction, sample or synthetic dataset for reproducibility when the original data could not be shared, explicit seed control, intentional type handling (e.g., typed function signatures), and archival codebases on persistent platforms.

Several limitations deserve consideration. First, this scoping review used an LLM-assisted approach to characterize articles and associated code repositories. Although our validation indicated good performance against human evaluation, automated classification is imperfect. Misclassification could modestly affect point estimates, and so these estimates should be interpreted as indicators of overall trends rather than exact counts. Second, our cohort was restricted to studies citing TRIPOD and TRIPOD+AI and retrievable through PubMed Central. This likely overrepresents studies attentive to reporting guidance and may therefore even overestimate code-sharing prevalence relative to broader prediction model literature.

Despite these caveats, this study provides the first large-scale, repository-level evaluation of code sharing practices in prediction model research aligned with TRIPOD guidance. The findings highlight that increasing availability is necessary but insufficient; improving reproducibility will require clearer expectations regarding documentation, dependency specification, licensing, and executable structure. These empirical observations directly inform the development of the TRIPOD-Code extension, which seeks to define minimal, feasible reporting standards for analytical code in clinical prediction studies.

%% file: Sections/5_conclusion.tex
\section{Conclusion}

Among prediction model studies citing TRIPOD or TRIPOD+AI, analytical code sharing remains limited. When code is made available, repository practices frequently lack features that support reproducibility and reuse. Although code availability has increased over time, direct examination of repositories suggests that availability alone does not ensure usability or computational reproducibility. Improving reproducibility in prediction model studies will require clearer expectations around documentation, dependency specification, licensing, and structure. These results provide an empirical foundation for the TRIPOD-Code extension and suggest that journals, funders, and reporting frameworks have a critical role in aligning researchers toward meaningful, reproducible code sharing. Attention must not only be given to whether code is shared, but to how it is shared.

%% file: Sections/6_others.tex
\section*{Contributors}

Conceptualisation: TS, RG, LAC, KGMM, GSC, CL, TP. Methodology, data curation, formal analysis, writing - original draft: TS, RG, TP. Software: TS, RG. Writing - review and editing: TS, RG, CAG, LB, HL, HCL, LAC, KGMM, GSC, CL, TP. Supervision: CL, TP. All authors had access to the data and had final responsibility for the decision to submit for publication.

\section*{Data and code availability statement}

The code used to conduct this scoping review, along with the datasets generated at every step of the analysis, is available in the following GitHub repository: \mbox{https://github.com/thomas-sounack/TRIPOD-Code}. The repository is archived and accessible via this link: https://doi.org/10.5281/zenodo.18897513.

\section*{Declaration of interests}

TP serves or has recently served on the editorial board of \textit{npj Scientific Data}, \textit{PLOS Digital Health}, and \textit{Artificial Intelligence in Medicine}. LAC is Editor in Chief of \textit{PLOS Digital Health}. GSC and KGMM are Editors in Chief of \textit{Diagnostic and Prognostic Research}. GSC is a statistical editor for the \textit{BMJ}. GSC and KGMM are the lead authors of the TRIPOD and the TRIPOD+AI reporting guidelines. HL is an associate editor of \textit{npj Digital Medicine}. HCL is guest editor of \textit{Artificial Intelligence in Medicine}. CAG serves on the editorial board of \textit{CHEST}.

\section*{Acknowledgements}

Open Access funding provided by the MIT Libraries. GSC is supported by the EPSRC (Engineering and Physical Sciences Research Council) grant for “Artificial intelligence innovation to accelerate health research” (EP/Y018516/1), and MRC-NIHR Better Methods Better Research grant (MR/Z503873/1). GSC was supported by the National Institute for Health and Care Research (NIHR) Birmingham Biomedical Research Centre. GSC is a National Institute for Health and Care Research (NIHR) Senior Investigator. The views expressed in this article are those of the author and not necessarily those of the NIHR or the Department of Health and Social Care. TP is supported by the National Institute of Health (NIH-R01EB030362, NIH-OT2OD032701, NIH-U24EB037545). TP, LAC, HC, and HL are supported by a grant of the Boston-Korea Innovative Research Project through the Korea Health Industry Development Institute (KHIDI), funded by the Ministry of Health \& Welfare, Republic of Korea (grant no.: RS-2024-00403047, NTIS no.: 2460003034). CAG is supported by NIH/NHLBI K23HL169815, a Parker B. Francis Opportunity Award, and an American Thoracic Society Unrestricted Grant.

%% file: Sections/99_appendix.tex
\section{PRISMA Checklist}\label{prisma-checklist}

\begin{figure}[H]
    \centering
    \includegraphics[width=0.9\linewidth]{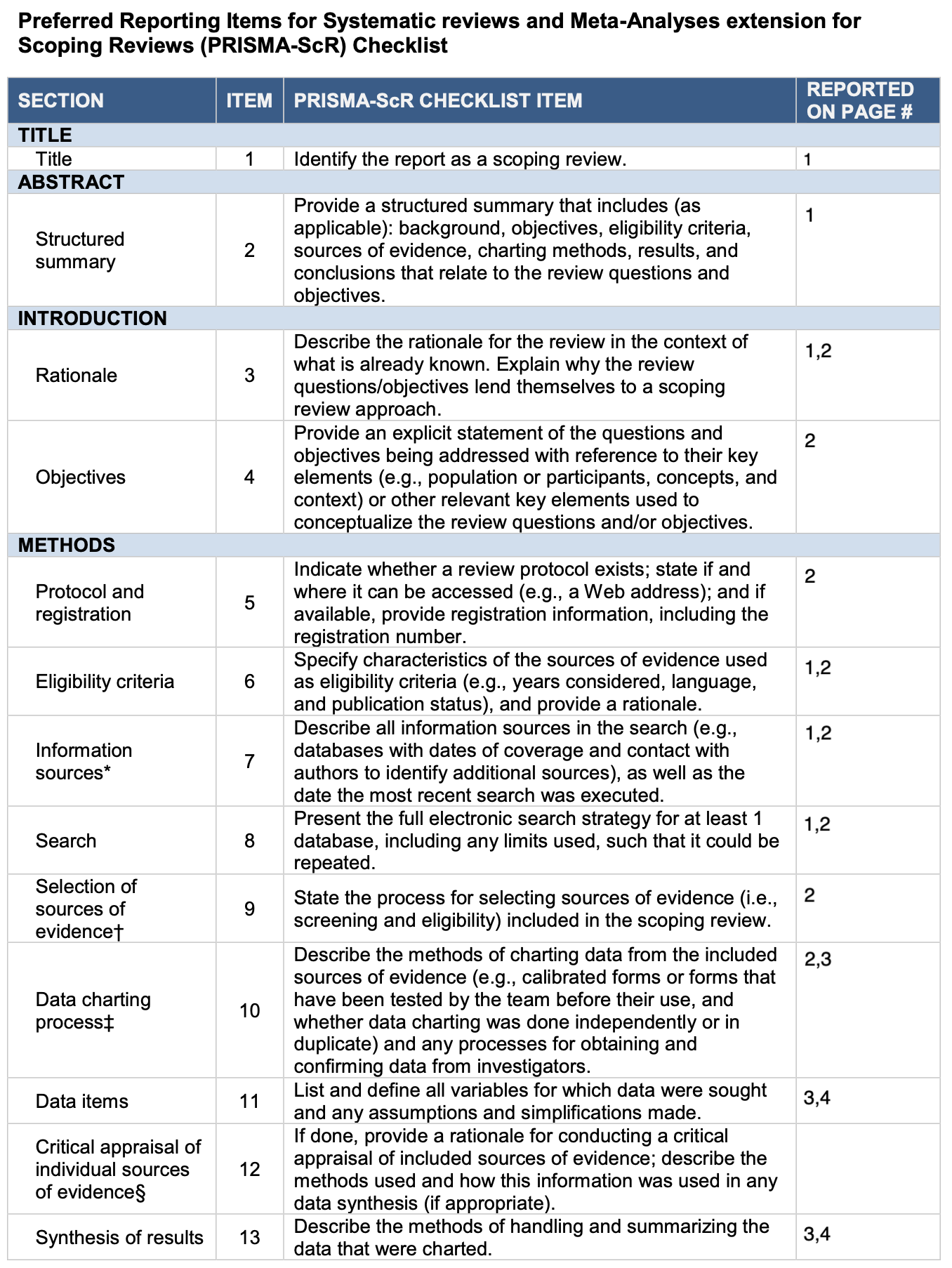}
\end{figure}

\begin{figure}[H]
    \centering
    \includegraphics[width=0.9\linewidth]{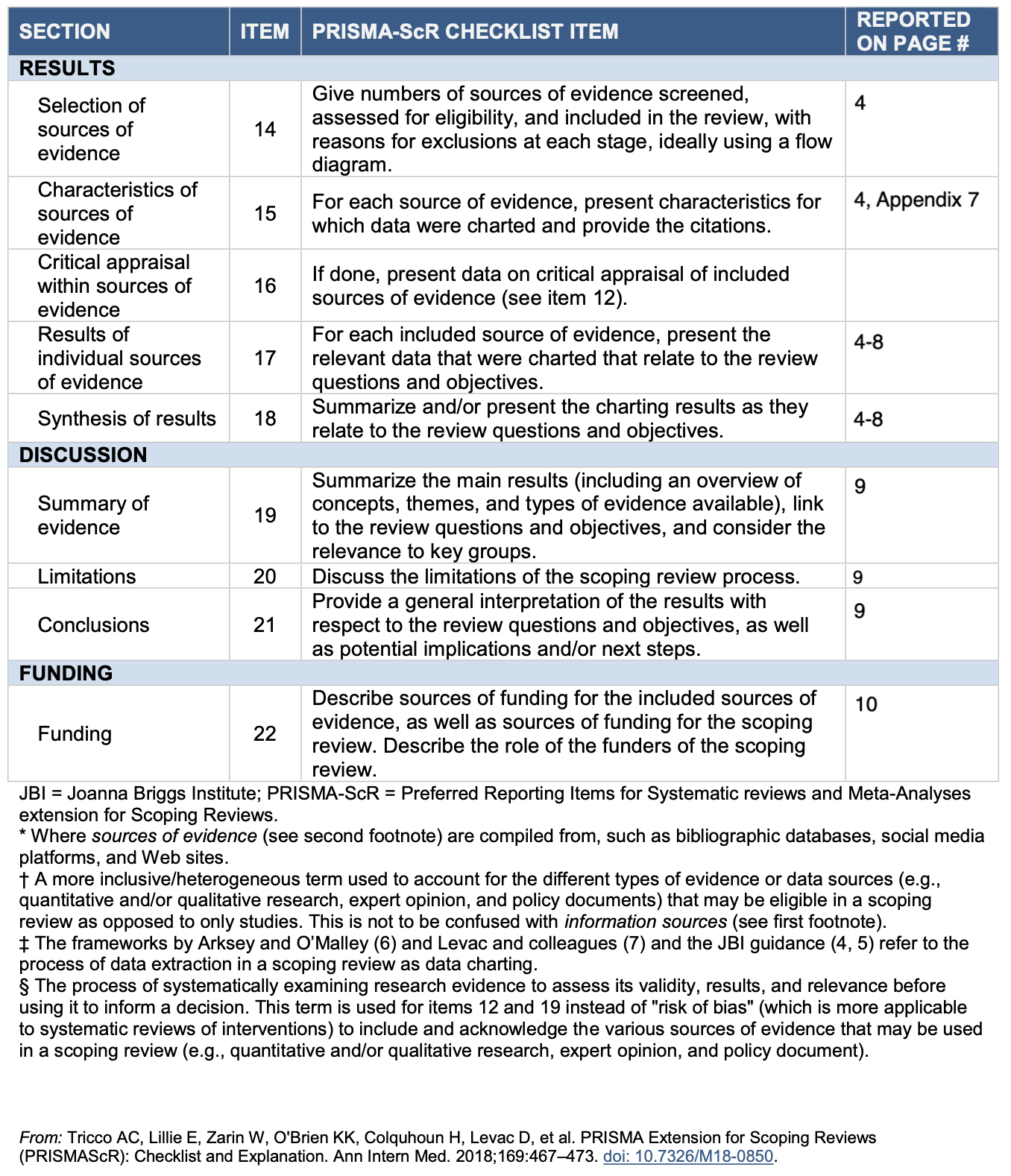}
\end{figure}

\newpage
\section{Article screening and metadata extraction prompt and output schema}\label{prompt-articles}

\subsection{Prompt}
\begin{quote}
We include a paper in our analysis if it is a project in which a multivariable prediction model is developed, updated or validated using any statistical or machine learning technique.
It has to be in the study itself, not as a reference to another paper. For instance, a protocol that details how a prediction model will be run in a future study does not qualify.
If the study itself uses any statistical model, such as a COX regression model, a multivariable logistic regression, for example, it should be included and will count as meeting the criteria. Any modality of prediction model is included, also include time series models, image-based models, and text-based models.
Given a paper, decide if the paper would fit this criteria. You need to provide a a boolean match (in the is\_match field) and a reason for whether the paper meets the criteria (in the reason field).
In the field country\_first\_author\_institution, return the country of origin based on the university / company of affiliation of the first author. Use the ISO 3166 standard name of the country in your response. If the information cannot be found, return 'not reported'.
Additionally, return a URL to the paper's code repository (in the field repo\_url) if it is provided and the paper is a match.
The repository should be reported to contain the code used to conduct the study, do not report a repository for a library or tool that was developed external to the paper but was used in the study.
Report only code repositories, not model or data repositories. If a user account link to a repository platform is reported instead of a repository, you can report it.
Otherwise, always report the root of the repository, ignoring releases or subfolders that could be included in the link.
If in the supplementary material or appendix section the code is reported, return 'Appendix' as the URL (but it has to explicitly mention that this supplemental contains the code). If a DOI is provided as the repository link, format it in a resolvable URL form.
In the field code\_statement\_locations, return the list of all locations in the paper where a code availability statement appears, if a repo\_url is found. Use ['other'] if the code availability statement location does not fit the available categories.
In the field code\_statement\_sentence, if repo\_url is found, return the sentence introducing the repository url (without the url itself), for example 'The code can be found here:', 'Our code is provided here'.
\end{quote}

\subsection{Output schema}
\begin{lstlisting}[language=Python,basicstyle=\small,breaklines=true]
CodeStatementLocation = Literal[
    "abstract",
    "introduction",
    "methods",
    "results",
    "discussion",
    "data_availability_section",
    "code_availability_section",
    "supplementary_material",
    "other",
]

class PaperAssessment(BaseModel):
    """
    Structured assessment of whether a paper meets inclusion criteria
    for multivariable prediction model studies, with justification and
    associated code repository information if applicable.
    """

    is_match: bool = Field(
        ...,
        description="Whether the paper meets the inclusion criteria for a multivariable prediction model study.",
    )
    reason: str = Field(
        ...,
        description="Brief explanation justifying why the paper does or does not meet the criteria.",
    )
    country_first_author_institution: str = Field(
        ...,
        description=(
            "The country of origin based on the affiliation of the first author. Use the ISO 3166 standard name of the country in your response."
            "Return 'not reported' if the information is not found"
        ),
    )
    repo_url: Optional[str] = Field(
        ...,
        description=(
            "URL to the paper's code repository if the paper is a match. "
            "Use 'Appendix' if code is explicitly stated to be in supplementary materials"
        ),
    )
    code_statement_locations: Optional[List[CodeStatementLocation]] = Field(
        ...,
        description=(
            "All locations in the paper where a code availability statement appears if a repo_url is found. "
            "Use ['other'] if the code availability statement location does not fit the available categories"
        ),
    )
    code_statement_sentence: Optional[str] = Field(
        ...,
        description="If repo_url is found, the sentence introducing the repository url (without the url itself), eg. 'The code can be found here:'",
    )
\end{lstlisting}

\newpage
\section{Code repository characterization prompt and output schema}\label{prompt-repositories}

\subsection{Prompt}
\begin{quote}
You will be provided the tree of a repository and its code. Use it to assess the quality of the repository.
You should return a boolean for each of these categories:
- is\_empty: is the repository empty? A repository is considered empty if it contains no files, only files that are empty, or only a README file.
- contains\_readme: does the repository contain instructions on how the code is structured and how to use it (such as a README.md, README.txt or README file)?
- readme\_purpose\_and\_outputs: (don't return anything for this field if contains\_readme is false) do these instructions provide an overview of the purpose of the code repository, and its expected outputs?
- contains\_requirements: does the repository specify the software dependencies used to run the code in a separate file (for example as a requirements.txt, environment.yml, or pyproject.toml file) or in the README file?
- requirements\_dependency\_versions: (don't return anything for this field if requirements\_dependency\_versions is false) does the requirements file specify dependency version requirements?
- contains\_license: does the repository include a license file specifying how others can use this code?
- sufficient\_code\_documentation: does the code include sufficient inline documentation or comments explaining the purpose and functionality of key components of the code for a user to understand its logic?
- is\_modular\_and\_structured: is the code organized into modular, reusable components using functions and classes where appropriate, rather than consisting of a single or a few long scripts?
- implements\_tests: does the repository include unit tests or functional tests to verify that the code works as intended? This may include test scripts, test files, or embedded assertions that check whether inputs and outputs behave as expected.
- fixes\_seed\_if\_stochastic: (If applicable, don't return anything if the repository doesn't use stochastic processes) if using stochastic processes (e.g., random number generation, machine learning models), is the repository setting fixed random seeds to ensure reproducibility?
- hardware\_requirements: are hardware requirements listed?
- contains\_link\_to\_paper: does the repository contain a link to the paper it was used for?
- contains\_citation: does the repository include a citation to the paper, in the format of a latex citation key or in plain text?
- includes\_data\_or\_sample: does the repository include either the original dataset or a sample dataset for demonstration purposes?
- comments\_and\_explanations: provide additional comments and explanations regarding the repository's quality, strengths, weaknesses, or any notable aspects that may not be fully captured by the boolean assessments above.
- coding\_languages: (if the repository contains code) list all programming languages used in the repository.
\end{quote}

\subsection{Output schema}

\begin{lstlisting}[language=Python,basicstyle=\small,breaklines=true]
class RepoAssessment(BaseModel):
    # Relevance
    is_empty: bool = Field(
        ...,
        description=(
            "Whether the repository is empty. Consider it empty if it contains no files, "
            "only empty files, or only a README file."
        ),
    )

    # README
    contains_readme: bool = Field(
        ...,
        description=(
            "Whether the repository contains usage/structure instructions (e.g., README.md/README.txt/README)."
        ),
    )
    readme_purpose_and_outputs: Optional[bool] = Field(
        ...,
        description=(
            "If contains_readme is True, whether the README provides an overview of the repository purpose "
            "and expected outputs. Do not return anything if contains_readme is False."
        ),
    )

    # Requirements
    contains_requirements: bool = Field(
        ...,
        description=(
            "Whether the repository specifies software dependencies either in a dedicated file "
            "(e.g., requirements.txt, environment.yml, pyproject.toml) or in the README."
        ),
    )
    requirements_dependency_versions: Optional[bool] = Field(
        ...,
        description=(
            "If contains_requirements is True, whether dependencies include version constraints "
            "(e.g., package==1.2.3, >=, ~=). Do not return anything if contains_requirements is False."
        ),
    )

    # License
    contains_license: bool = Field(
        ...,
        description="Whether the repository includes a license file describing usage permissions.",
    )

    # Documentation
    sufficient_code_documentation: bool = Field(
        ...,
        description=(
            "Whether the code contains sufficient inline comments/docstrings explaining key components "
            "so a user can understand the logic."
        ),
    )

    # Modularity
    is_modular_and_structured: bool = Field(
        ...,
        description=(
            "Whether code is organized into modular, reusable components (functions/classes/modules) "
            "rather than a few long scripts."
        ),
    )

    # Testing
    implements_tests: bool = Field(
        ...,
        description=(
            "Whether the repository includes tests (unit/functional), test files/scripts, or meaningful "
            "assertions verifying expected behavior."
        ),
    )

    # Reproducibility
    fixes_seed_if_stochastic: Optional[bool] = Field(
        ...,
        description=(
            "If the repository uses stochastic processes (e.g., random sampling, ML training), whether it "
            "sets fixed random seeds for reproducibility. Do not return anything if stochasticity is not applicable."
        ),
    )
    lists_hardware_requirements: bool = Field(
        ...,
        description="Whether hardware requirements (e.g., GPU/CPU/RAM) are stated anywhere in the repository.",
    )

    # Citation and Linking
    contains_link_to_paper: bool = Field(
        ...,
        description="Whether the repository includes a link (URL/DOI/arXiv/PubMed) to the associated paper.",
    )
    contains_citation: bool = Field(
        ...,
        description=(
            "Whether the repository provides a citation for the paper (e.g., plain text citation, BibTeX entry, "
            "CITATION.cff, or a LaTeX citation key)."
        ),
    )

    # Data
    includes_data_or_sample: bool = Field(
        ...,
        description=(
            "Whether the repository includes the original dataset or a sample/demo dataset sufficient to run "
            "or demonstrate the code."
        ),
    )

    # Free-text notes
    comments_and_explanations: Optional[str] = Field(
        ...,
        description=(
            "Additional comments about repository quality, strengths/weaknesses, and notable aspects not fully "
            "captured by the boolean fields."
        ),
    )

    # Languages
    coding_languages: Optional[List[str]] = Field(
        ...,
        description=(
            "If the repository contains code, return all programming languages used. In a list"
            "For example, ['python', 'r', 'sql']."
            "Do not return anything if there is no code in the repository."
        ),
    )
\end{lstlisting}

\section{Annotation codebook}\label{annotation-codebook}

\subsection*{Article annotation guidelines}

\subsubsection*{Background}

This document provides detailed instructions for annotators involved in the TRIPOD-Code project. The goal of this annotation task is to evaluate the availability and quality of code repositories linked to studies that develop, update, or validate multivariable prediction models. Annotations will help assess the transparency, reproducibility, and reusability of code associated with these studies. The guidelines outlined below are designed to ensure consistency across annotators and support downstream analysis. Unless otherwise specified, fields should be annotated using binary values: True, False, or left blank when not applicable.

\subsubsection*{Task}

You will be working with a spreadsheet in which each row corresponds to a scientific paper that cites the TRIPOD or TRIPOD+AI reporting guidelines, based on a citation search conducted up to August 11th, 2025. For each paper, your task is to assess whether it falls within the scope of the TRIPOD-Code project and to annotate relevant metadata, particularly regarding the presence and quality of any associated code repositories.

\subsubsection*{Determining whether a paper should be considered}

We include a paper in our analysis if it reports on the development, updating, or validation of a multivariable prediction model, regardless of whether it uses traditional statistical methods or machine learning techniques. The model must involve multiple predictors and aim to estimate the probability of an outcome for individual instances. This determination is guided by the following scope statement from our protocol: ``the focus of TRIPOD-Code is on reports of projects in which a multivariable prediction model is developed, updated or validated using any statistical or machine learning technique''

\subsubsection*{Data preparation}

To support reproducibility, the PDF files or web pages used during annotation will be saved to shared storage before the annotation process begins.

In the annotation spreadsheet, annotators will see a field labeled Annotation Support Document, which links directly to the saved version stored on shared storage. An additional column will list the original URL used to retrieve the document, for reference.

\subsection*{Repository annotation guidelines}

\subsubsection*{Code presence and relevance}

\paragraph{Relevance}

Is empty: is the repository empty? A repository is considered empty if it contains no files, only files that are empty, or only a README file.

A repository is also considered empty if the link is inaccessible.

\subsubsection*{Code quality and reproducibility}

\paragraph{README}

Is included: does the repository contain instructions on how the code is structured and how to use it (such as a README.md, README.txt or README file)?

Overview purpose and outputs: do these instructions provide an overview of the purpose of the code repository, and its expected outputs?

\paragraph{Requirements}

Is included: does the repository specify the software dependencies used to run the code in a separate file (for example as a requirements.txt, environment.yml, or pyproject.toml file) or in the README file?

Dependency versions: does the requirements file specify dependency version requirements?

\paragraph{License}

Is included: does the repository include a license file specifying how others can use this code?

\paragraph{Documentation}

Documentation: does the code include sufficient inline documentation or comments explaining the purpose and functionality of key components of the code for a user to understand its logic?

\paragraph{Modularity}

Modularity and Structure: is the code organized into modular, reusable components using functions and classes where appropriate, rather than consisting of a single or a few long scripts?

\paragraph{Testing}

Unit or Functional testing: does the repository include unit tests or functional tests to verify that the code works as intended? This may include test scripts, test files, or embedded assertions that check whether inputs and outputs behave as expected.

\paragraph{Reproducibility}

Seed: if using stochastic processes (e.g., random number generation, machine learning models), is the repository setting fixed random seeds to ensure reproducibility?

Hardware requirements: are hardware requirements listed?

\subsubsection*{Sharing and long term availability}

\paragraph{Citation and Linking}

Repo includes link to paper: link to preprint / publisher website

Repo includes citation file / text: can be in the format of a citation key for latex or in plain text.

\paragraph{Data}

Sample for demo: does the repository include either the original dataset or a sample dataset for demonstration purposes? The goal is to get a simple overview of the required input format as well as the type of each field.

\newpage
\section{Article screening: evaluation metrics and error analysis}\label{eval-articles}

The table below summarizes the model’s performance in classifying articles according to the selection criteria, using the 500 articles annotated by TS and RG.

\begin{table}[H]
\centering
\label{tab:scope-classification-performance}
\begin{tabular}{lcccc}
\toprule
 & Precision & Recall & F1 Score & Support \\
\midrule
Out of scope & 0.93 & 0.94 & 0.93 & 105 \\
In scope & 0.98 & 0.98 & 0.98 & 395 \\
\textbf{Weighted average} & \textbf{0.97} & \textbf{0.97} & \textbf{0.97} & \\
\bottomrule
\end{tabular}
\end{table}

Additionally, we evaluated the LLM’s ability to extract repository links from full-text articles. In the same set of 500 articles, annotators identified 39 references to code. The model recovered 36 of these correctly, returned an incorrect link for one article, and returned no link for two articles. The model also flagged code references in four articles where annotators had not recorded any.

On manual review, nearly all discrepancies reflected issues with the human labels rather than model errors. For the two articles where the model returned no link \cite{choiDevelopmentValidationMachine2023, yuPredictingMetabolicDysfunction2025}, the annotator-identified links were not code repositories (a Shiny application in one case and a CDC dataset in the other). In one article \cite{tranAssessingMachineLearning2024}, the authors provided a GitHub repository containing checkpoints of prediction models (https://github.com/Tran031194/abpmML); annotators did not count it as a repository because it was not strictly analysis code, but the classification was inherently ambiguous. In another case \cite{ottenhoffPredictingMortalityIndividual2021}, the repository was provided as a DOI rather than a direct URL (https://doi.org/10.5281/zenodo.4077342), whereas the model returned the equivalent Zenodo landing page (https://zenodo.org/record/4077342). Finally, for three articles \cite{habibClaimsBasedMachineLearning2025, zhangDistinctImmunologicalSignatures2025, panditAnalyzingHistoricalFuture2022}, the model correctly identified that code was provided in supplementary materials, which annotators had missed.

\section{Repository characterization: evaluation metrics and error analysis}\label{eval-repos}

Of the 39 code references recorded during article-level annotation, four pointed to an appendix rather than an external repository. The remaining 35 were annotated by TS and RG. Six links were classified as empty or inaccessible by our pipeline. Among these, three were genuinely inaccessible repositories and were flagged as such by the repository utilities. Two links did not in fact point to repositories and therefore could not be accessed or processed by the utilities. The remaining link pointed to an empty repository, which the LLM correctly flagged as empty. Model performance on the remaining 29 repositories is reported in the table below.

\begin{table}[H]
\centering
\begin{tabular}{lcccc}
\toprule
 & Precision & Recall & F1 Score & Support \\
\midrule
Contains README & 1.00 & 1.00 & 1.00 & 29 \\
README purpose and outputs & 0.70 & 1.00 & 0.82 & 24 \\
Contains Requirements & 0.62 & 1.00 & 0.76 & 29 \\
Requirements dependency versions & 1.00 & 0.86 & 0.92 & 8 \\
Contains license & 1.00 & 1.00 & 1.00 & 29 \\
Sufficient Code Documentation & 0.71 & 0.71 & 0.71 & 29 \\
Is modular and structured & 1.00 & 0.85 & 0.92 & 29 \\
Implements tests & 1.00 & 1.00 & 1.00 & 29 \\
Fixes seed if stochastic & 0.71 & 0.83 & 0.77 & 22 \\
Lists hardware requirements & 0.50 & 1.00 & 0.67 & 29 \\
Contains link to paper & 0.50 & 1.00 & 0.67 & 29 \\
Contains citation & 1.00 & 0.57 & 0.73 & 29 \\
Includes data or sample data & 0.71 & 1.00 & 0.83 & 29 \\
\textbf{Weighted average} & \textbf{0.80} & \textbf{0.91} & \textbf{0.83} & \\
\bottomrule
\end{tabular}
\end{table}

Overall, the model offered strong performance on objective criteria, such as whether the repository contains a README or a license file. We also note a high recall performance overall, indicating that true positives were rarely missed and that errors would tend to overestimate code sharing practices, making our conclusions conservative. However, on items that are more subjective, such as whether documentation was sufficient or whether a README file adequately described the purpose and outputs of a repository, the model scored lower against human labels. These being subjective criteria, annotators found them difficult to apply consistently, reflecting ambiguity in the criteria definition rather than straightforward model error. That ambiguity itself is informative and underscores the need for clearer guidelines defining what is sufficient documentation and reporting in code repositories.

\section{Articles selected}\label{csv-articles-selected}

Due to the size of our review cohort, we cannot individually cite all 3,967 articles included in this scoping review. However, the set of PMIDs can be found in \href{https://github.com/thomas-sounack/TRIPOD-Code/blob/main/data/02_paper_assessment/paper_assessment_pred.parquet.br}{this file of our GitHub repository}, with every row where `is\_match` is true being included in our review cohort.

\section{Code sharing sentences}\label{code-sharing-sentences}

The table below lists the code sharing sentences that occurred more than once in the 522 articles reporting code, with the corresponding number of occurrences. The search was case sensitive. We did not observe a standard code sharing statement, as most of them (94.4\%) only occurred once.

\begin{table}[H]
\centering
\begin{tabular}{p{0.7\textwidth}c}
\toprule
Code Sharing Sentence & Number Of Occurrences \\
\midrule
The complete code used for model development and evaluation in this project is publicly available on GitHub: & 8 \\
The source code for preprocessing and analyzing the data is available on GitHub & 3 \\
The code to build and train the model is openly available on github. & 2 \\
The code used to conduct the analysis is available on GitHub & 2 \\
The implementation of the deep learning framework is available at: & 2 \\
Code is available at & 2 \\
The code base for training the deep learning models used in this study is available at: & 2 \\
All code, metadata and documentation for this project is publicly available at & 2 \\
We have made the code and data used in this study available on a public repository. & 2 \\
The code developed for constructing the algorithms along with the original dataset, is available on Github & 2 \\
All code developed and used throughout this study has been made open source and is available on GitHub. & 2 \\
\bottomrule
\end{tabular}
\end{table}

\newpage
\section{Distribution of languages used in the repositories}\label{language-repo-distribution}

The following figure shows the distribution of all the languages identified by the LLM in the set of repositories.

\begin{figure}[H]
    \centering
    \includegraphics[width=0.9\linewidth]{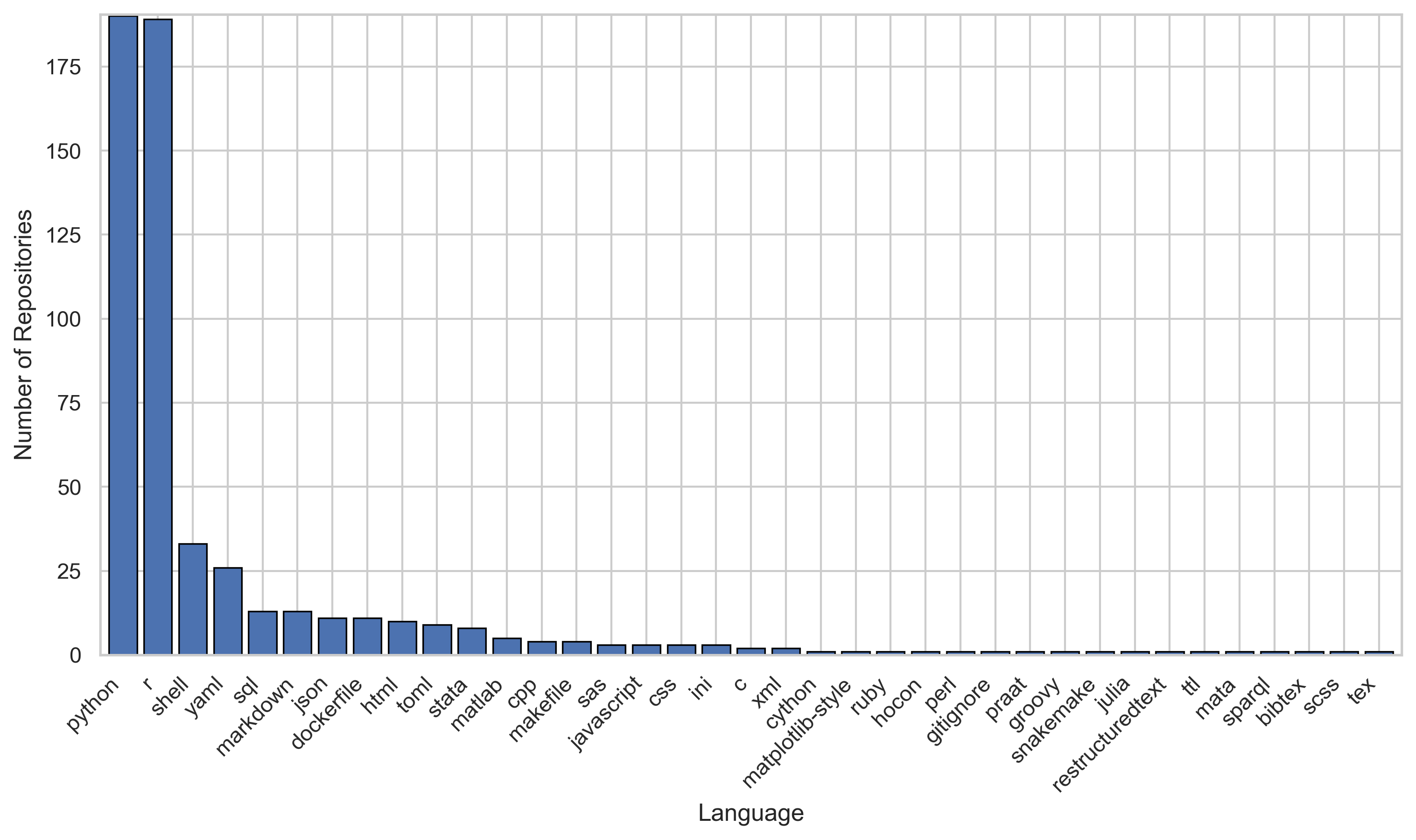}
\end{figure}